\documentclass[letter-paper, 10 pt, conference]{ieeeconf}  

\IEEEoverridecommandlockouts                              

\usepackage{graphics} 
\usepackage{epsfig} 
\usepackage{mathptmx} 
\usepackage{times} 
\usepackage{amsmath} 
\usepackage{amssymb}  
\usepackage{hyperref}
\usepackage{subcaption}
\usepackage[usenames,dvipsnames]{xcolor, colortbl}
\definecolor{Gray}{gray}{0.9}
\usepackage{algpseudocode,algorithm}
\usepackage{tikz}
\usetikzlibrary{arrows,positioning,scopes}

\title{\LARGE \bf Design and Selection of Additional Residuals to Enhance Fault Isolation of a Turbocharged Spark Ignited Engine System*}

\author{Kok Yew Ng$^{1,3}$, Erik Frisk$^{2}$, and Mattias Krysander$^{2}$
\thanks{*This research was supported by Volvo Car Corporation in Gothenburg, Sweden.}
\thanks{$^{1}$K. Y. Ng is with the Engineering Research Institute, Ulster University, Newtownabbey, BT37 0QB, UK.
        Emails: {\tt\footnotesize mark.ng@ulster.ac.uk}}%
        \thanks{$^{2}$E. Frisk and M. Krysander are with the Department of Electrical Engineering, Link\"oping University, Link\"oping 58183, Sweden.
        Emails: {\tt\footnotesize\{erik.frisk, mattias.krysander\}@liu.se}}
\thanks{$^{3}$K. Y. Ng is also affiliated with the School of Engineering, Monash University, 47500 Selangor, Malaysia.
       Email: {\tt\footnotesize kok.yew.ng@monash.edu}}%
}

\begin{document}

\pagestyle{empty}
\onecolumn
\noindent {\large \copyright ~2020 IEEE.  Personal use of this material is permitted.  Permission from IEEE must be obtained for all other uses, in any current or future media, including reprinting/republishing this material for advertising or promotional purposes, creating new collective works, for resale or redistribution to servers or lists, or reuse of any copyrighted component of this work in other works. \\~\\

\noindent This is a peer-reviewed and accepted version of the following in press document.} \\~\\

\noindent {\large K. Y. Ng, E. Frisk, and M. Krysander. ``Design and Selection of Additional Residuals to Enhance Fault Isolation of a Turbocharged Spark Ignited Engine System,'' {\it 7th International Conference on Control, Decision and Information Technologies (CoDIT'20) (In Press)}, 2020.}

\twocolumn
\newpage

\maketitle
\thispagestyle{empty}
\pagestyle{empty}

\begin{abstract}
This paper presents a method to enhance fault isolation without adding physical sensors on a turbocharged spark ignited petrol engine system by designing additional residuals from an initial observer-based residuals setup. The best candidates from all potential additional residuals are selected using the concept of sequential residual generation to ensure best fault isolation performance for the least number of additional residuals required. A simulation testbed is used to generate realistic engine data for the design of the additional residuals and the fault isolation performance is verified using structural analysis method.
\end{abstract}


\section{INTRODUCTION}
Fault diagnosis of dynamic systems has always been an interesting and exciting area of research, especially with the advancements in automation and manufacturing \cite{Chen2012, Isermann2016}. It is crucial for these autonomous systems, be it robotic systems in a manufacturing plant or a self-driving vehicle, to know the health of the systems such that corrective measures can be carried out in the event of a failure. As such, a fault diagnosis scheme must be able to perform two main tasks: fault detection, i.e. the ability to determine if a fault is present in the system, and fault isolation, i.e. to locate the root cause of the fault \cite{Ding2008}.

These fault diagnosis schemes are usually designed using either hardware redundancy or analytical redundancy methods. The hardware redundancy method utilizes multiple identical sensors to measure the same variable of the system. A voting mechanism is then applied to determine the presence of a fault \cite{Radu}.
One of the main challenges of diagnostic systems is to improve fault isolation without adding physical sensors in order to reduce weight and the overall cost of the system. As a result, many of modern diagnostic systems are designed using model-based or analytical redundancy methods. These methods usually use observers, which are constructed using the mathematical equations describing the system dynamics, to estimate states of the system \cite{Gao, Ng2012}. Using the same control input to drive both observer and the actual system, the difference between the actual outputs of the system and the estimated outputs of the observer is computed to produce the residuals, which are then processed to perform fault diagnosis \cite{edwards2000sliding}. One of the limitations to these techniques is that the number of sensors available would affect the quality of the diagnosis, i.e. more sensors (and hence, residuals) would lead to better fault isolation performance \cite{FRISK2009364}.

This paper proposes to use the concept of sequential residual generation reported in \cite{Svard} to design and select additional residuals for a vehicular turbocharged spark ignited engine system with data obtained using the simulation testbed in \cite{CSM}. The purpose is to improve fault isolation without adding physical sensors onto the engine system.

This paper is organized as follows: Section \ref{sec:Engine} introduces the engine system and the problem statement; Section \ref{section:Res} presents the design and generation of residuals using a conventional model-based method; Section \ref{section:Addres} provides some backgrounds on the design and generation of additional residuals; Section \ref{sec:EngineApp} shows the application of additional residuals on the engine system, which includes simulation results and discussions; and Section \ref{sec:Conclusion} provides some conclusions.


\section{PROBLEM STATEMENT} \label{sec:Engine}
This paper addresses the issue of fault isolation in a vehicular turbocharged spark ignited (TCSI) engine system.
The engine system has 13 states, six actuators and seven measured outputs. See Table \ref{tab:states}.
The specifications and parameters of the reference engine system can be found in \cite{CSM}.

This research considers 11 faults of interest located in various subsystems of the engine system, where six are variable faults,
one actuator fault,
 and four sensor measurement faults.
 Table \ref{tab:fault} shows the faults of interest and their descriptions. Only single fault scenarios are considered in this paper.
\begin{table}[t!]
\caption{\label{tab:states}The states, actuators, and sensors (measured states) of the engine system and their descriptions.}
\centering
\begin{tabular}{lll} \hline
{\bf Variable}	& {\bf Description}   & {\bf Unit} \\ \hline
{\bf States} & & \\
$T_{af}$    & Temperature at the air filter   & K \\
$p_{af}$    & Pressure at the air filter      & Pa \\
$T_{c}$     & Temperature at the compressor {\bf (measured)}   & K \\
$p_{c}$     & Pressure at the compressor {\bf (measured)}      & Pa \\
$T_{ic}$    & Temperature at the intercooler {\bf (measured)}  & K \\
$p_{ic}$    & Pressure at the intercooler {\bf (measured)}     & Pa \\
$T_{im}$    & Temperature at the intake manifold {\bf (measured)} & K \\
$p_{im}$    & Pressure at the intake manifold {\bf (measured)} & Pa \\
$T_{em}$    & Temperature at the exhaust manifold & K \\
$p_{em}$    & Pressure at the exhaust manifold & Pa \\
$T_{t}$     & Temperature at the turbine      & K \\
$p_{t}$     & Pressure at the turbine         & Pa \\
$W_{af}$    & Mass flow at the air filter {\bf (measured)}   & kg/s \vspace{2mm} \\
{\bf Actuators} & & \\
$A_{th}$    & Throttle position area          & m$^2$ \\
$u_{wg}$    & Wastegate input                 & [0...1] \\
$\omega_{eREF}$ & Reference engine speed      & rad/s \\
$\lambda$   & Air-fuel ratio                  & [--] \\
$T_{amb}$   & Ambient Temperature             & K \\
$p_{amb}$   & Ambient pressure                & Pa \vspace{2mm} \\ \hline
\end{tabular}
\end{table}

\begin{table}[t!]
\caption{\label{tab:fault}Faults of interest and their descriptions.}
\centering
\begin{tabular}{cl} \hline
{\bf Fault}			& {\bf Description} \\ \hline
$f_{p_{af}}$		& Loss of pressure in the air filter 	 					\\
$f_{C_{vol}}$ 		& Intake valve timing stuck at arbitrary position 			\\
$f_{W_{af}}$		& Air leakage between the air filter and the compressor 	\\
$f_{W_{c}}$			& Air leakage between the compressor and the intercooler 	\\
$f_{W_{ic}}$		& Air leakage between the intercooler and the throttle 		\\
$f_{W_{th}}$ 		& Air leakage after the throttle in the intake manifold		\\
$f_{x_{th}}$ 		& Throttle position actuator error							\\
$f_{y_{W_{af}}}$ 	& Air filter flow sensor fault								\\
$f_{y_{p_{im}}}$	& Intake manifold pressure sensor fault 					\\
$f_{y_{p_{ic}}}$ 	& Intercooler pressure sensor fault 						\\
$f_{y_{T_{ic}}}$ 	& Intercooler temperature sensor fault 						\\ \hline
\end{tabular}
\end{table}

Given that the engine system is highly nonlinear with many interconnected subsystems, diagnostic systems usually monitor multiple components simultaneously, although they are quite independent from each other. As a result, a fault that is present in the engine system can trigger several monitors or manifest into other types of faults, hence affecting fault isolation performance of the diagnostic systems \cite{goodloe2010monitoring}. This is critical as the ability to identify and isolate the root fault from the manifested faults enables the replacement of the correct faulty components to enhance the reliability of the overall system \cite{Scacchioli}. This also helps to ensure the safety of the occupants onboard the vehicle as well as other road users.

The goal is to design and generate additional useful residuals that would be useful to improve fault isolation without adding physical sensors to the systems.


\section{DESIGN AND GENERATION OF RESIDUALS} \label{section:Res}
\begin{figure}[t!]
  \centering
  \includegraphics[width=\columnwidth]{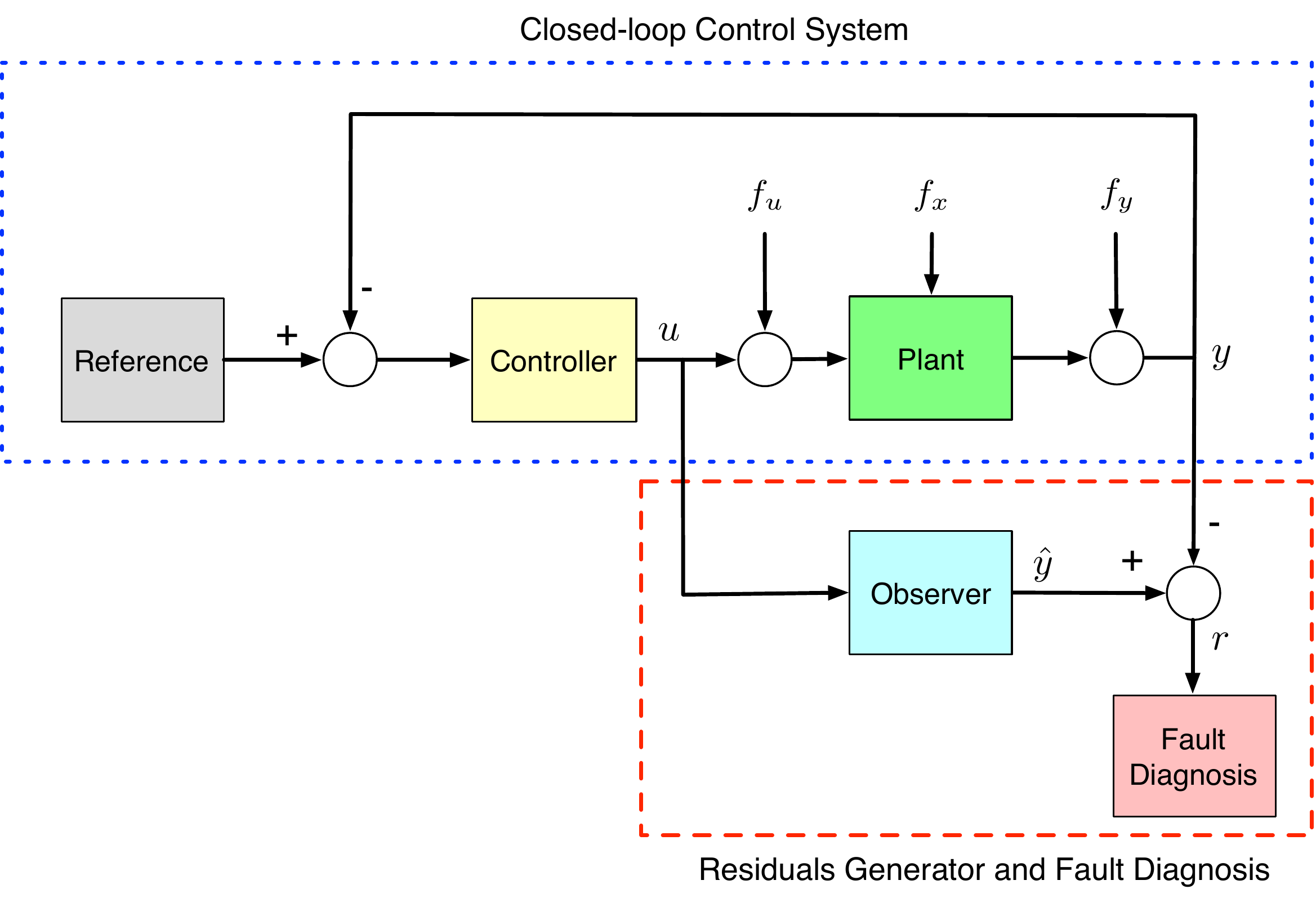}
  \caption{Block diagram representation of the closed-loop control system and the residuals generator.}
  \label{fig:FTC}
\end{figure}

Fig. \ref{fig:FTC} shows a typical block diagram of the closed-loop feedback system with the residuals generator. The blocks within the blue dotted box form the closed-loop control system while the blocks within the red dashed box form the residuals generator and the fault diagnosis scheme. Both plant and observer are driven by the same control input, where the plant would then produce the output $y$ while the observer would produce an estimate of the output $\hat y$. The difference between the actual output and its estimate is then used to generate the residual for fault detection, i.e. $r = \hat y - y$. During nominal fault-free scenario, the residual $r$ would have a mean of zero. In the presence of a fault, the residual $r$ would have a nonzero mean and hence, a fault has been detected.
\begin{table}[t!]
  \caption{Default ``Original 7'' residuals generated for fault detection based on the sensors setup of the engine system.}
  \centering
  \begin{tabular}{cl}
    \hline
    {\bf Residual}	&	{\bf Description} \\
    \hline
    $r_{T_{c}}$		& 	Residual for Compressor Temperature Sensor \\
    $r_{p_{c}}$		& 	Residual for Compressor Pressure Sensor \\
    $r_{T_{ic}}$	& 	Residual for Intercooler Temperature Sensor \\
    $r_{p_{ic}}$	& 	Residual for Intercooler Pressure Sensor \\
    $r_{T_{im}}$	& 	Residual for Intake Manifold Temperature Sensor \\
    $r_{p_{im}}$	& 	Residual for Intake Manifold Pressure Sensor \\
    $r_{W_{af}}$	& 	Residual for Air Filter Mass Flow Sensor \\
    \hline
  \end{tabular}
  \label{tab:residuals}
\end{table}%

\begin{figure}[t!]
  \centering
  \includegraphics[height=2in,width=\columnwidth]{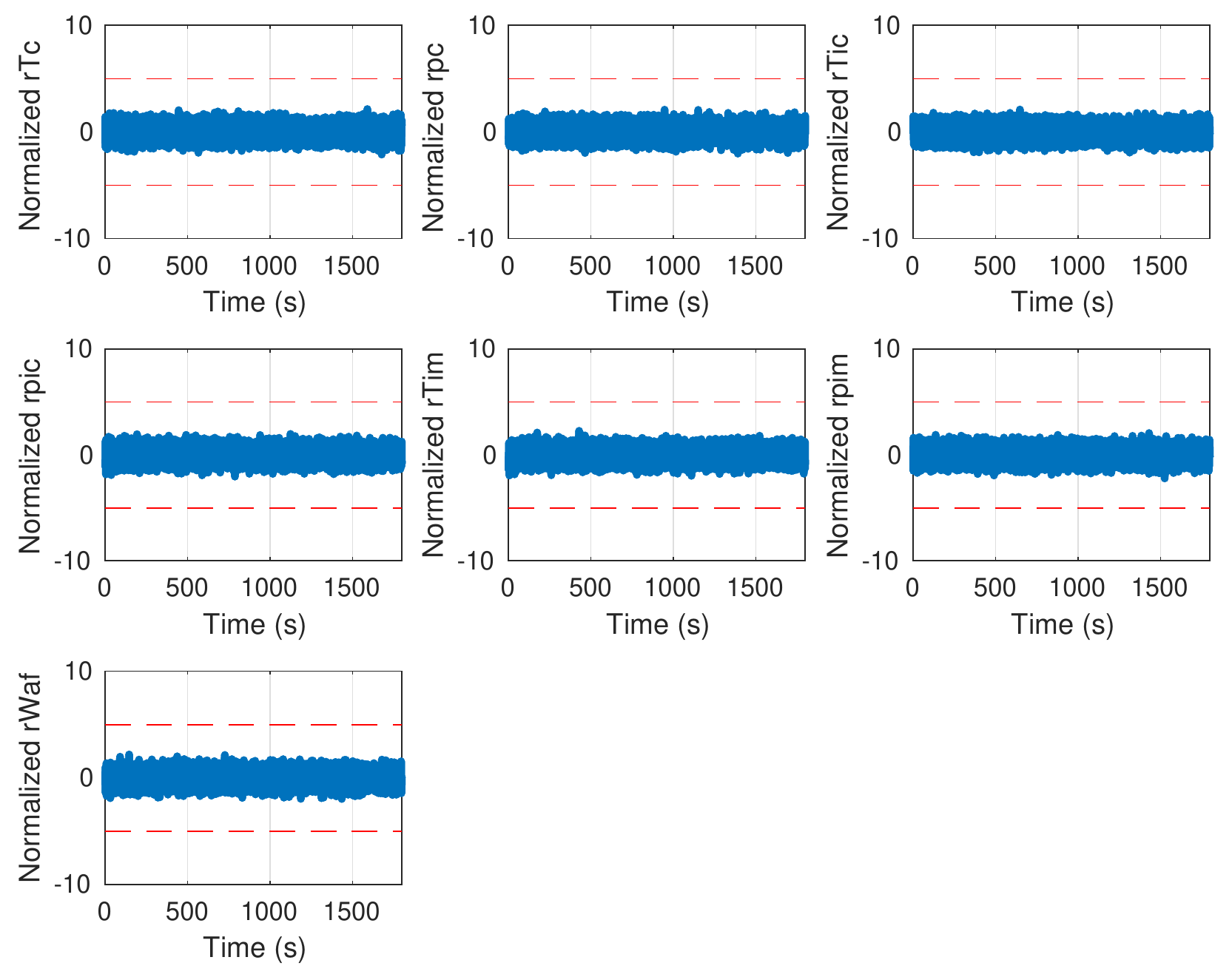}
  \caption{Normalized plots of the ``Original 7'' for a fault-free scenario. The dashed lines are the fault detection thresholds.}
  \label{fig:faultfreedef}
\end{figure}

\begin{figure}[t!]
  \centering
  \includegraphics[height=2in,width=\columnwidth]{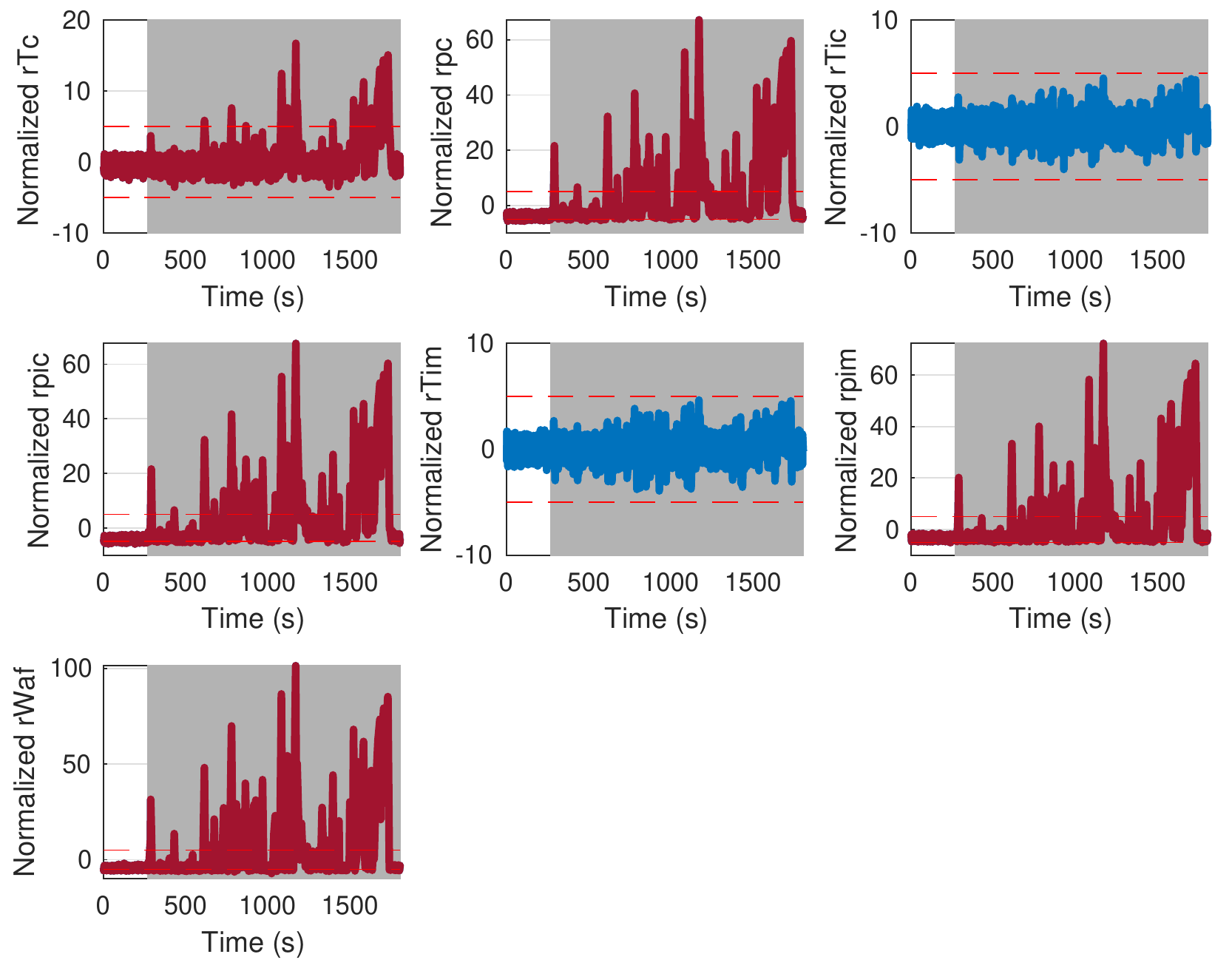}
  \caption{Normalized plots of the ``Original 7'' for a $f_{p_{af}}$ fault. The dashed lines are the fault detection thresholds. The plots in red are the residuals that have triggered. The shaded regions show the duration when the fault is active.}
  \label{fig:fpafdef}
\end{figure}

As shown in Section \ref{sec:Engine} and Table \ref{tab:states}, the reference engine system has seven measured outputs.
As such, there are seven residuals that can be generated. These residuals are called the ``Original 7'' (see Table \ref{tab:residuals}). Fig. \ref{fig:faultfreedef} shows the ``Original 7'' for a nominal fault-free scenario during the Worldwide Harmonized Light Vehicles Test Procedure (WLTP) run using the simulation testbed while Fig. \ref{fig:fpafdef} shows the ``Original 7'' for a $f_{p_{af}}$ fault during the same driving cycle. The horizontal red dashed lines are the thresholds set for fault detection. A fault is said to have been detected when one or more residuals exceed the thresholds, i.e. $|r| > J$. In this research, the thresholds were tuned using nominal fault-free data to achieve a tradeoff between false detection and missed detection rates. Therefore, the thresholds were initially set at $J = 5$. The gray regions in Fig. \ref{fig:fpafdef} show the the duration when the fault was active. Fig. \ref{fig:fpafdef} also shows that $f_{p_{af}}$ have triggered five of the ``Original 7'' residuals: $r_{T_{c}}, r_{p_{c}}, r_{p_{ic}}, r_{p_{im}}$, and $r_{W_{af}}$ (the plots with red lines). The simulation was repeated with the other faults of interest in Table \ref{tab:fault} and the states of the residuals were recorded in a fault sensitivity matrix (FSM) as shown using the unshaded rows in Table \ref{tab:FSMdef}, i.e. residuals that are sensitive to a particular fault and have triggered are given a state of ``1'', and ``0'' otherwise.

\newcolumntype{g}{>{\columncolor{Gray}}c}
\newcolumntype{f}{>{\columncolor{Gray}}l}
\begin{table*}[t!]
  \caption{The FSM for the ``Original 7'' residuals (unshaded rows) and the additional residuals (shaded rows).}
  \centering
  \begin{tabular}{fggggggggggg}
    \hline
    \rowcolor{white}
    {\bf Residuals} 	&	{\bf $f_{p_{af}}$}	&	{\bf $f_{C_{vol}}$}	&	{\bf $f_{W_{af}}$}	&	{\bf $f_{W_{c}}$}	&	{\bf $f_{W_{ic}}$}	&	{\bf $f_{W_{th}}$}	&	{\bf $f_{x_{th}}$}	&	{\bf $f_{y_{p_{ic}}}$}	&	{\bf $f_{y_{p_{im}}}$}	&	{\bf $f_{y_{T_{ic}}}$}	&	{\bf $f_{y_{W_{af}}}$} \\
    \hline
    \rowcolor{white}
    $r_{T_{c}}$ 	& 1	& 1 & 0	& 1	& 0	& 0	& 0	& 0	& 0	& 0 & 0 \\
    \rowcolor{white}
    $r_{p_{c}}$		& 1	& 1	& 0	& 1	& 1	& 1	& 0	& 0	& 0 & 0	& 0 \\
    \rowcolor{white}
    $r_{T_{ic}}$	& 0	& 1	& 0	& 1	& 0	& 0	& 0	& 0	& 0 & 1	& 0 \\
    \rowcolor{white}
    $r_{p_{ic}}$	& 1	& 1	& 0	& 1	& 1	& 1	& 0	& 1	& 0	& 0	& 0 \\
    \rowcolor{white}
    $r_{T_{im}}$	& 0	& 1	& 0	& 1	& 0	& 0	& 0	& 0	& 0 & 0 & 0 \\
    \rowcolor{white}
    $r_{p_{im}}$	& 1	& 1	& 0	& 1	& 1	& 1	& 0	& 0	& 1 & 0	& 0 \\
    \rowcolor{white}
    $r_{W_{af}}$	& 1	& 1	& 1	& 1 & 1	& 1	& 0	& 0	& 0	& 0 & 1 \\ \hline
    $p_{im}$\_$_{W_{af}}$        	& 0    & 1    & 0    & 1   & 0    & 1    & 0     & 0     & 1     & 0     & 1 \\
    $T_{im}$\_$_{W_{af}}$        	& 0    & 1    & 0    & 1   & 0    & 1    & 0     & 0     & 0     & 0     & 0 \\
    $p_{ic}$\_$_{W_{af}}$        	& 0    & 1    & 0    & 1   & 0    & 1    & 1     & 1     & 0     & 0     & 1 \\
    $T_{ic}$\_$_{W_{af}}$        	& 0    & 1    & 0    & 1   & 0    & 1    & 0     & 0     & 0     & 1     & 0 \\
    $W_{af}$\_$_{p_{im}}$        	& 1    & 1    & 1    & 1   & 1    & 1    & 1     & 0     & 1     & 0     & 1 \\
    $T_{im}$\_$_{p_{im}}$        	& 1    & 1    & 0    & 1   & 1    & 0    & 0     & 0     & 1     & 0     & 0 \\
    $p_{ic}$\_$_{p_{im}}$        	& 1    & 1    & 0    & 1   & 1    & 1    & 0     & 1     & 1     & 0     & 0 \\
    $T_{ic}$\_$_{p_{im}}$        	& 0    & 1    & 0    & 0   & 0    & 0    & 0     & 0     & 0     & 1     & 0 \\
    $W_{af}$\_$_{T_{im}}$        	& 1    & 1    & 1    & 1   & 1    & 1    & 0     & 0     & 0     & 0     & 1 \\
    $W_{af}$\_$_{T_{ic}p_{ic}T_{im}p_{im}}$ & 1    & 1    & 1    & 1   & 1    & 1    & 1     & 1     & 1     & 0     & 1 \\
    $W_{af}$\_$_{p_{c}T_{im}p_{im}}$     & 1    & 1    & 1    & 1   & 1    & 1    & 1     & 1     & 1     & 0     & 1 \\
    $W_{af}$\_$_{p_{c}T_{ic}p_{ic}}$     & 1    & 1    & 1    & 1   & 1    & 1    & 1     & 1     & 0     & 1     & 1 \\
    $p_{ic}$\_$_{T_{c}p_{c}T_{im}p_{im}}$   & 0    & 0    & 0    & 0   & 1    & 0    & 0     & 1     & 1     & 0     & 0 \\
    $T_{ic}$\_$_{T_{c}p_{c}T_{im}p_{im}}$   & 0    & 0    & 0    & 0   & 1    & 0    & 0     & 0     & 1     & 1     & 0 \\ \hline
  \end{tabular}
  \label{tab:FSMdef}
\end{table*}%

\begin{figure}[t!]
  \centering
  \includegraphics[width=0.8\columnwidth]{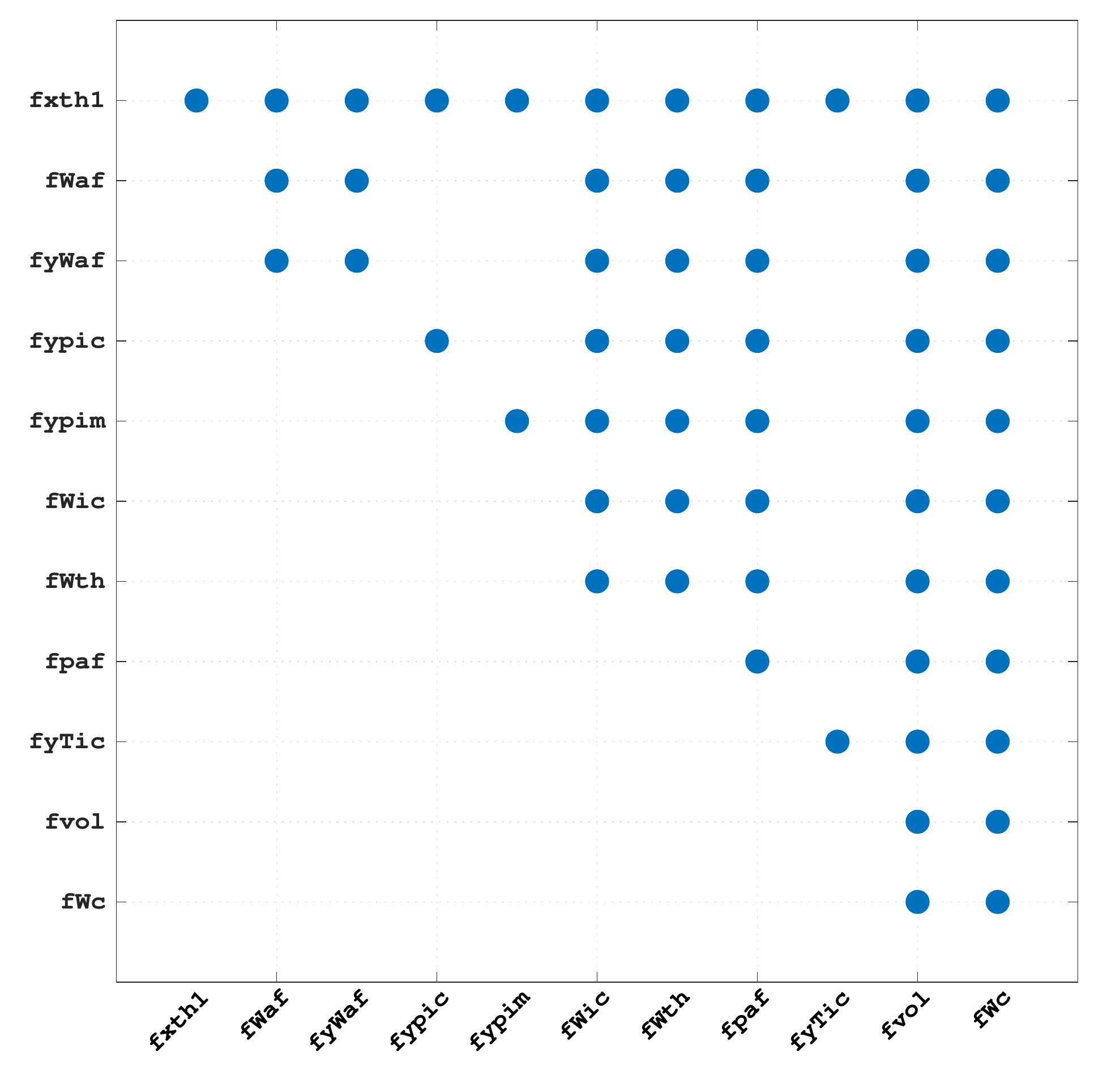}
  \caption{The FIM for the FSM of the ``Original 7'' in Table \ref{tab:FSMdef} (unshaded rows only).}
  \label{fig:FIMdef}
\end{figure}

Using data from the unshaded rows in Table \ref{tab:FSMdef}, a fault isolation matrix (FIM) could then be generated to analyse the fault isolation performance of using only the ``Original 7'' residuals (see Fig. \ref{fig:FIMdef}). Fig. \ref{fig:FIMdef} shows that only $f_{x_{th}}$ could be isolated from the other faults. As such, additional residuals can be designed to improve fault isolation. See \cite{Damadics} for some background studies on structural model, FSM, and FIM.

\section{DESIGN AND GENERATION OF ADDITIONAL RESIDUALS}\label{section:Addres}
With reference to \cite{Svard}, this section presents some backgrounds on the design of the additional residuals.

First, let's assume a system can be described using the following differential equations
\begin{eqnarray}
  \dot x_{1} &=& -x_{1} + u + f_{u}, \label{eq:x1dot} \\
  \dot x_{2} &=& x_{1} - x_{2}, \label{eq:x2dot}\\
  y_{1} &=& x_{1} + f_{1}, \label{eq:y1} \\
  y_{2} &=& x_{2} + f_{2}, \label{eq:y2}
\end{eqnarray}
where $x_{i}$ are the states, $u$ the measurable input, $y_{i}$ the measurable outputs, $f_{u}$ the input fault, and $\{f_{1},f_{2}\}$ are the output faults.

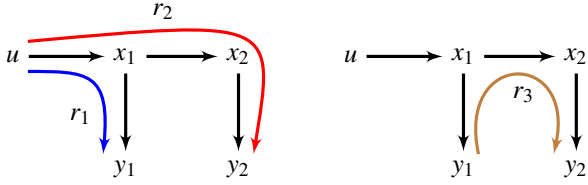
\begin{figure}[t!]
  \begin{subfigure}[t]{0.48\columnwidth}
  	\centering

\begin{tikzpicture}[auto, node distance=1.5cm, >=latex']
	\node (u) {$u$};
	\node (x1) [right of=u] {$x_{1}$};
	\node (x2) [right of=x1] {$x_{2}$};
	\node (y1) [below of=x1] {$y_{1}$};
	\node (y2) [below of=x2] {$y_{2}$};
	
	\draw[->, very thick] (u) -- (x1);
	\draw[->, very thick] (x1) -- (x2);
	\draw[->, very thick] (x1) -- (y1);
	\draw[->, very thick] (x2) -- (y2);
	
	\draw[->, blue, very thick] (0.2,-0.2) .. controls (1,-0.2) and (1.3,-0.2) .. (1.2,-1.3);
	\node at (0.9,-0.8) (r1) {$r_{1}$};
	\draw[->, red, very thick] (0.2,0.2) .. controls (3.5,0.5) and (3.5,0.5) .. (3.2,-1.3);
	\node at (2,0.6) (r2) {$r_{2}$};
	
%
%
\end{tikzpicture}
	\caption{Tracing from $u$ to $\{y_{1},y_{2}\}$ through $\{x_{1},x_{2}\}$ to generate $\{r_{1},r_{2}\}$.}
  	\label{fig:r1r2}
  \end{subfigure} ~
  \begin{subfigure}[t]{0.48\columnwidth}
  	\centering
	\begin{tikzpicture}[auto, node distance=1.5cm, >=latex']
	\node (u) {$u$};
	\node (x1) [right of=u] {$x_{1}$};
	\node (x2) [right of=x1] {$x_{2}$};
	\node (y1) [below of=x1] {$y_{1}$};
	\node (y2) [below of=x2] {$y_{2}$};
	
	\draw[->, very thick] (u) -- (x1);
	\draw[->, very thick] (x1) -- (x2);
	\draw[->, very thick] (x1) -- (y1);
	\draw[->, very thick] (x2) -- (y2);
	
	\draw[->, brown, very thick] (1.7,-1.3) .. controls (1.4,0.1) and (3,0.1) .. (2.7,-1.3);
	\node at (2.3,-.5) (r3) {$r_{3}$};
	\draw[->, white, very thick] (0.2,0.2) .. controls (3.5,0.5) and (3.5,0.5) .. (3.2,-1.3);

%
%
\end{tikzpicture}
	\caption{A third and new residual, $r_{3}$, which is used to estimate $y_{2}$, is generated by tracing $y_{1}$ to $y_{2}$.}
  	\label{fig:r3}
  \end{subfigure}
  \caption{Design of residuals by tracing paths from other measurable variables to the variables to be estimated.}
\end{figure}

The relationship between the input $u$ and the outputs $\{y_{1},y_{2}\}$ for the system in (\ref{eq:x1dot})--(\ref{eq:y2}) can be traced through the states $\{x_{1},x_{2}\}$, as shown in Fig. \ref{fig:r1r2}. Since there are 2 outputs, i.e. $y_{1}$ and $y_{2}$, an observer can be formulated to produce 2 residual signals, $r_{1}$ and $r_{2}$, respectively. Fig. \ref{fig:r1r2} also shows that each residual is generated by taking different routes from $u$ to $y_{1}$ and $y_{2}$, respectively. As such, residual $r_{1}$ can be designed to estimate $y_{1}$ using
\begin{eqnarray}
  \dot {\hat {x}}_{1} &=& -\hat x_{1} + u, \label{eq:x1dothat} \\
  \hat y_{1} &=& \hat x_{1}, \\
  r_{1} &=& y_{1} - \hat y_{1}, \label{eq:r1}
\end{eqnarray}
and residual $r_{2}$ can be designed to estimate $y_{2}$ using
\begin{eqnarray}
  \dot {\hat {x}}_{2} &=& \hat x_{1} - \hat x_{2}, \label{eq:x2dothat} \\
  \hat y_{2} &=& \hat x_{2}, \\
  r_{2} &=& y_{2} - \hat y_{2}, \label{eq:r2}
\end{eqnarray}
where $\hat x_{i}$ and $\hat y_{i}$ are the estimates of the states $x_{i}$ and the outputs $y_{i}$, respectively.

\begin{figure}[t!]
	\begin{subfigure}[t]{\columnwidth}
		\centering
  		\includegraphics[height=3cm,width=\columnwidth]{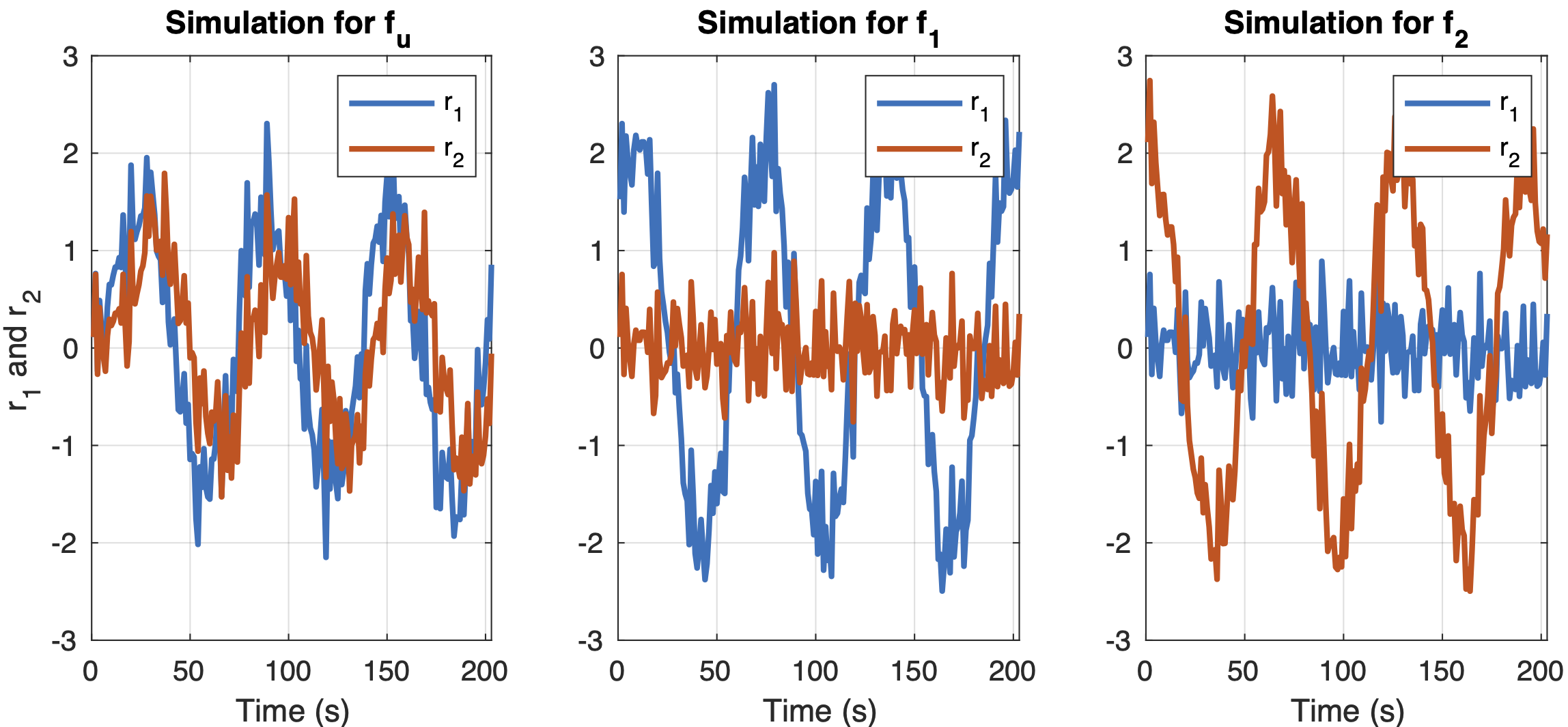}
  		\caption{Sensitivity of only the original residuals $\{r_{1},r_{2}\}$.}
  		\label{fig:Sim1} \vspace{3mm}
	\end{subfigure}
	\begin{subfigure}[t]{\columnwidth}
  		\centering
  		\includegraphics[height=3cm,width=\columnwidth]{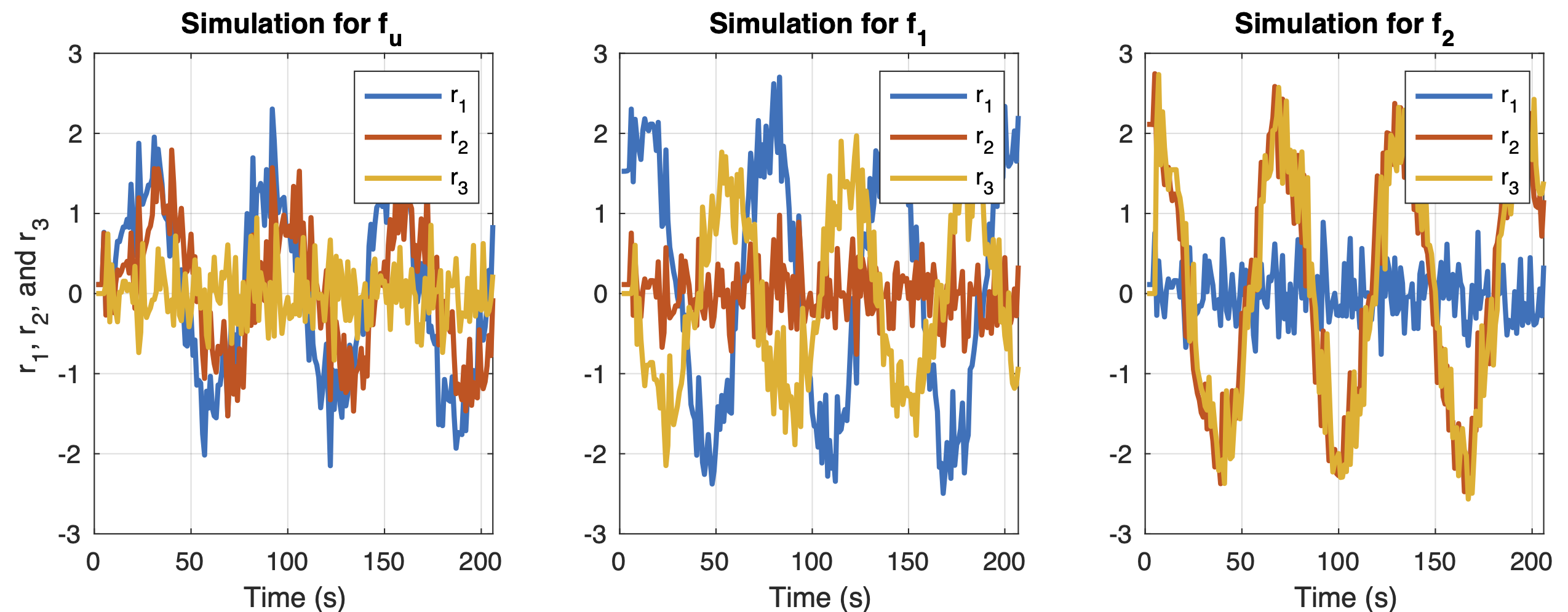}
  		\caption{Sensitivity of the original and additional residuals $\{r_{1},r_{2},r_{3}\}$.}
  		\label{fig:Sim2}
	\end{subfigure}
	\caption{Simulations showing the performance of the original residuals $\{r_{1},r_{2}\}$ and then with the additional residual $r_{3}$ towards the faults in system (\ref{eq:x1dot})--(\ref{eq:y2}).}
\end{figure}

Simulating this system with the residual generators in Matlab/Simulink where the faults are injected as sinusoidal signals; $f_{u} = 2sin(t)$, $f_{1} = 2sin(t + \frac{\pi}{4})$, and $f_{2} = 2sin(t + \frac{\pi}{2})$, Fig. \ref{fig:Sim1} shows that residual $r_{1}$ is sensitive to $f_{1}$ and $f_{u}$ while $r_{2}$ is sensitive to $f_{2}$ and $f_{u}$. As a result, the FSM for $\{r_{1}$, $r_{2}\}$ can be represented by the unshaded rows in Table \ref{tab:FSMr1r2}. The unshaded rows in Table \ref{tab:FSMr1r2} show that the sensor faults can be successfully isolated from each other. However, the fault sensitivity does not allow for the elimination of $f_{u}$ from the sets of diagnosis, i.e. the detection of either $f_{1}$ or $f_{2}$ also includes a detection of $f_{u}$. Therefore, a new residual has to be designed to achieve better fault isolation.
\newcolumntype{g}{>{\columncolor{Gray}}c}
\newcolumntype{f}{>{\columncolor{Gray}}l}
\begin{table}[t!]
  \caption{The FSM for $\{r_{1},r_{2}\}$ (unshaded rows), and the additional residual $r_{3}$ (shaded row).}
  \centering
  \begin{tabular}{gggg}
    \hline
    \rowcolor{white}
    {\bf Residual} 	&	$f_{1}$	&	$f_{2}$	&	$f_{u}$ \\
    \hline
    \rowcolor{white}
    $r_{1}$ 	& 1 & 0	& 1 \\
    \rowcolor{white}
    $r_{2}$		& 0	& 1	& 1	\\ \hline
    $r_{3}$		& 1	& 1	& 0	\\ \hline
  \end{tabular}
  \label{tab:FSMr1r2}
\end{table}

Let's reconsider the system (\ref{eq:x1dot})--(\ref{eq:y2}). Assume now that there exists another route for the estimation of $y_{2}$. Removing the equation for $y_{1}$ in (\ref{eq:y1}) and using $\hat x_{1}$ to estimate $y_{1}$ as shown in (\ref{eq:x1hat}), a new path can be traced to $y_{2}$ (see Fig. \ref{fig:r3}), which results in the generation of a third residual, $r_{3}$ in (\ref{eq:r3}).
\begin{eqnarray}
  \hat x_{1} &=& y_{1}, \label{eq:x1hat} \\
  \dot {\hat x}_{2} &=& \hat x_{1} - \hat x_{2}, \\
  \hat y_{2} &=& \hat x_{2}, \\
  r_{3} &=& y_{2} - \hat y_{2}. \label{eq:r3}
\end{eqnarray}

\begin{figure}[t!]
  \centering
  \includegraphics[width=0.5\columnwidth]{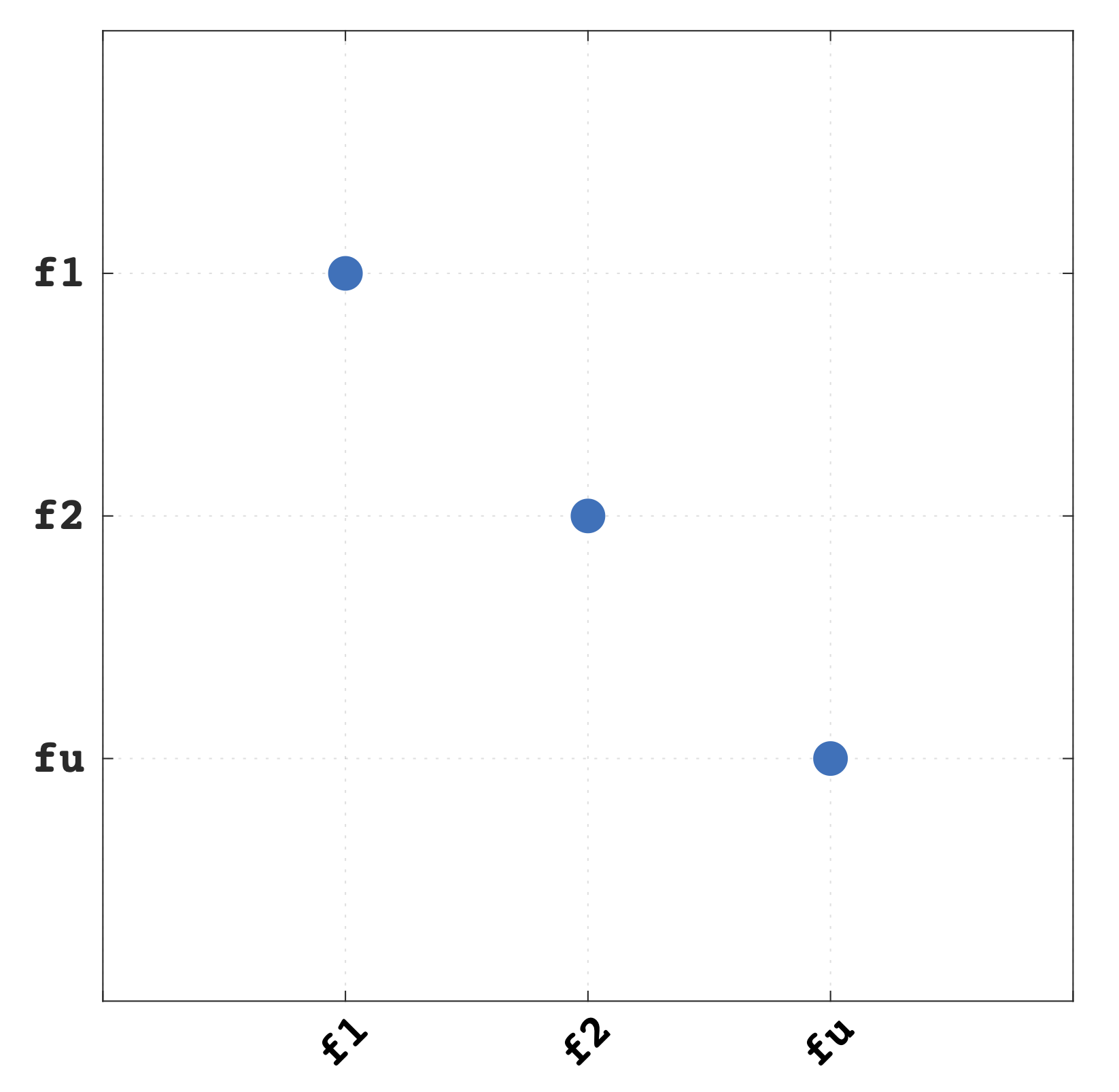}
  \caption{The FIM with the inclusion of $r_{3}$.}
  \label{fig:FIM}
\end{figure}

The omission of $u$ from this path removes the sensitivity of $r_{3}$ towards $f_{u}$. As a result, the FSM for the residuals can be updated to include the shaded row in Table \ref{tab:FSMr1r2}. The updated FSM shows that $f_{u}$ can be isolated from the sensor faults using the new residual and this is reflected in the FIM in Fig. \ref{fig:FIM}. Table \ref{tab:resDiag} shows the relationships between the triggered residuals and the diagnosis decisions.

\begin{table}[t!]
  \caption{The triggered residuals and the diagnosis decisions for the isolated faults.}
  \centering
  \begin{tabular}{cc}
    \hline
    \rowcolor{white}
    {\bf Triggered Residuals} 	&	{\bf Diagnosis Decision} \\
    \hline
    $r_{1}$ and $r_{2}$ 	& $f_{u}$ \\
    $r_{1}$ and $r_{3}$		& $f_{1}$ \\
    $r_{2}$	and $r_{3}$		& $f_{2}$ \\ \hline
  \end{tabular}
  \label{tab:resDiag}
\end{table}

Another possible additional residual, $r_{4}$, can be generated by removing the equation for $y_{2}$ in (\ref{eq:y2}) and using $\hat x_{2}$ to estimate $y_{2}$. Hence, a path can be traced to $y_{1}$ (the opposite direction of $r_{3}$) to provide an estimate of $y_{1}$. However, it is not utilized in this example as all faults can already be isolated from each other with the the original residuals and only one additional residual, $r_{3}$.

\section{APPLICATION TO A TCSI ENGINE SYSTEM} \label{sec:EngineApp}
For the TCSI engine, it is desired to have more residuals than the ``Original 7'' to make the fault isolation investigation relevant. As such, the procedure discussed in Section \ref{section:Addres} is applied to design additional residuals for the engine model. However, due to the higher number of original residuals used by the engine (7 residuals) compared to the ones used in the example (2 residuals) in Section \ref{section:Addres}, there are more routes/paths that can be chosen to generate the additional residuals. For example, a system with $n$ original residuals, there are $(n \times 2^{n-1}) - n$ possible candidates as additional residuals. Therefore, the TCSI engine system with the ``Original 7'' residuals has 441 potential candidates as additional residuals. Some of these additional residuals would enhance fault isolation performance while the remaining ones, although being able to provide estimation of the measurable variables, do not necessarily improve fault isolation. To reiterate, for the example presented in Section \ref{section:Addres}, $r_{3}$ is sufficient to enhance fault isolation performance while $r_{4}$ is not necessary although it would still provide an estimate of $y_{1}$. Therefore, the challenge is to determine the minimum number of additional residuals that would improve the fault isolation performance most. This is critical as the requirements for computing power increase with the number of additional residuals generated.

A Matlab-based algorithm is used to generate all potential candidates for additional residuals, as well as to search for the best candidates and to determine the least number of additional residuals that can guarantee the best fault isolation performance. 

Algorithm \ref{alg:genRes} first explains the procedure to generate all possible additional residuals from the current residuals setup. Given that the TCSI engine system has 7 original residuals ($n = 7$), there are $(2^{n} - 1)$ combinations in residuals setup using the variable {\tt cycle}, where `{\tt ON}' and `{\tt OFF}' states of these residuals are represented in binary. The main loop is then used to generate all $(n \times 2^{n-1}) - n$ possible candidates by sequentially removing the corresponding equations and rows from the relations matrix of the structural model. Algorithm \ref{alg:autoChoice} would then utilize the results from Algorithm \ref{alg:genRes} to select the minimum best candidates for additional residuals to enhance fault isolation performance. This is done by running all original residuals and potential candidates through an optimization function {\tt autoSelect()}, where each iteration would choose a potential candidate as one of the best candidates and update the overall FIM of all residuals. The algorithm would terminate when the optimization function returns no best candidate.

The resultant FIM for the original residuals together with this new set of best additional candidates would provide the updated analysis on fault isolation performance.
\renewcommand{\algorithmicrequire}{\textbf{Input:}}
\renewcommand{\algorithmicensure}{\textbf{Output:}}
\begin{algorithm}[t!]
  \caption{Generation of all possible additional residuals from the ``Original 7''.
    \label{alg:genRes}}
    \begin{algorithmic}[1]
	    \Require $n, StructuralModel, FSM$

		\State $cycle \gets binary( [1:( 2^{n}-1 )] )$ \Comment{All residuals \\ \hfill combinations}

	    \For{$i \gets 1$ \textbf{to} $(2^{n} - 1)$}	\Comment{Main loop}
	    	\State $sumSens \gets getSumSensON( cycle( i ) )$
			\State $posSens \gets getPosSensON( cycle( i ) )$
			\State $nameSens \gets getNameSensON( posSens )$
			\State $tempRelMat \gets getRelMat( StructModel )$

			\For{$j \gets 1$ \textbf{to} $sumSens$}	\Comment{Remove corresponding \\ \hfill eqs. from relations matrix}
				\State $eqNo \gets getEqNo( StructModel, nameSens )$
				\State \textbf{remove} $tempRelMat( eqNo )$
			\EndFor

			\State $genNewRes( tempRelMat )$
	    \EndFor

	    \Ensure $allAddRes, allAddResFSM$
    \end{algorithmic}
\end{algorithm}

\begin{algorithm}[t!]
  \caption{Selection of minimum additional residuals from all possible additional residuals to enhance fault isolation.
    \label{alg:autoChoice}}
    \begin{algorithmic}[1]
	    \Require $n, nFaults, allAddResFSM$

	    \State $bestFIM \gets ones( nFaults )$
	    \Repeat
	    	\State $selRes \gets autoSelect( bestFIM, allAddResFSM )$
		    \State $bestAddRes \gets [bestAddRes; selRes]$
		    \State $bestFIM \gets FSMtoFIM( allAddResFSM( bestRes ) )$
	    \Until{$isEmpty( selRes )$}

	    \Ensure $bestAddRes$

	    \\
	    \Function{autoSelect}{ FIM, FSM }

			\For{$i \gets 1$ \textbf{to} $( n \times 2^{n-1}) $}
				\State $zerosIdx \gets getPosZeros( FSM( i ) )$
				\State $onesIdx \gets getPosOnes( FSM( i ) )$
				\If{$sum( FIM( onesIdx, zerosIdx ) ) > 0$}
					\State $selRes \gets i$
				\Else
					\State $selRes \gets [~]$
				\EndIf
			\EndFor

			\State \Return $selRes$
	    \EndFunction
    \end{algorithmic}
\end{algorithm}

\subsection{RESULTS AND DISCUSSIONS}
\begin{figure}[t!]
  \centering
  \includegraphics[height=2.4in,width=\columnwidth]{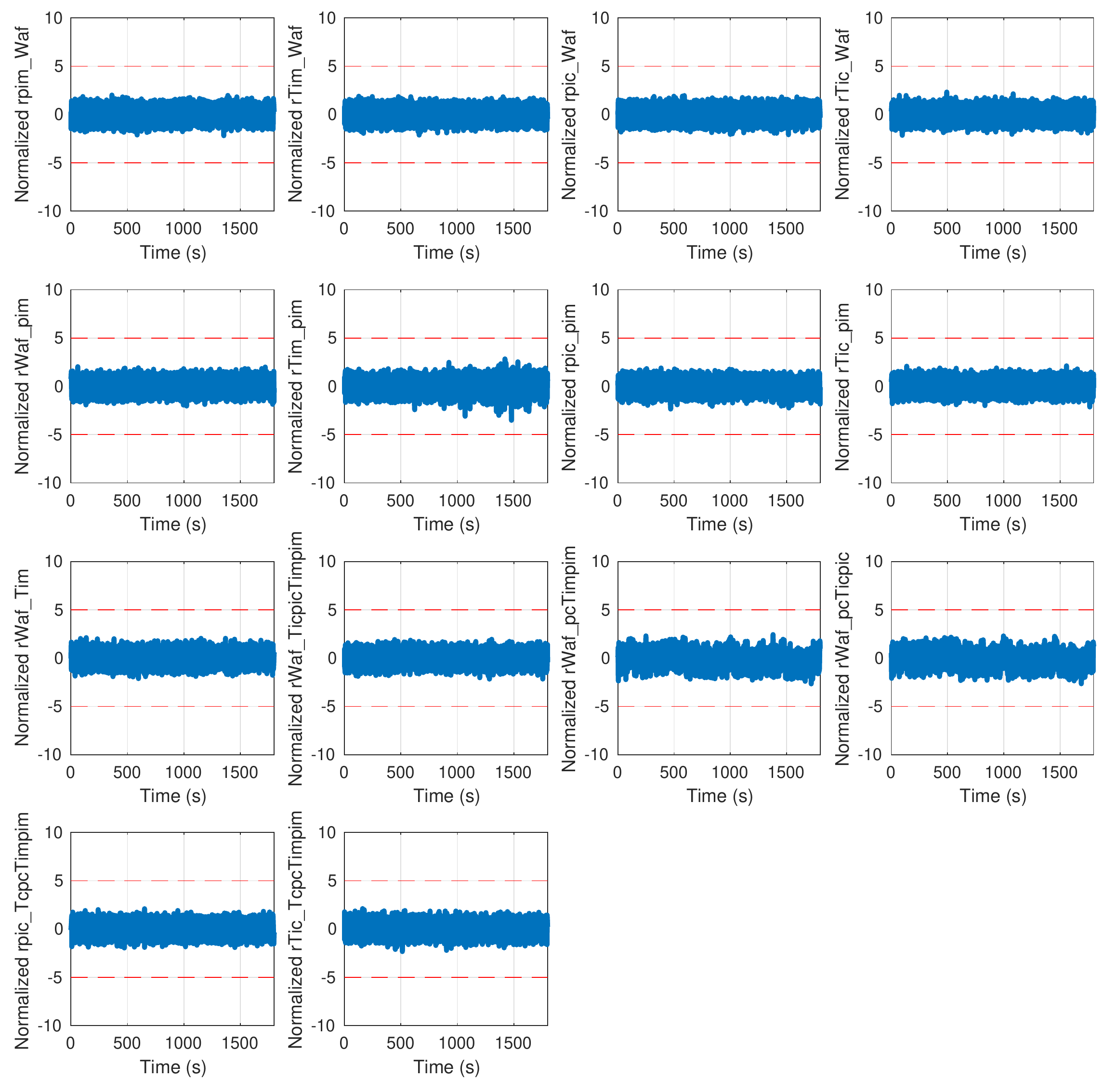}
  \caption{Normalized plots of the 14 additional residuals for a fault-free scenario. The dashed lines are the fault detection thresholds.}
  \label{fig:faultfreeadd}
\end{figure}

\begin{figure}[t!]
  \centering
  \includegraphics[height=2.4in,width=\columnwidth]{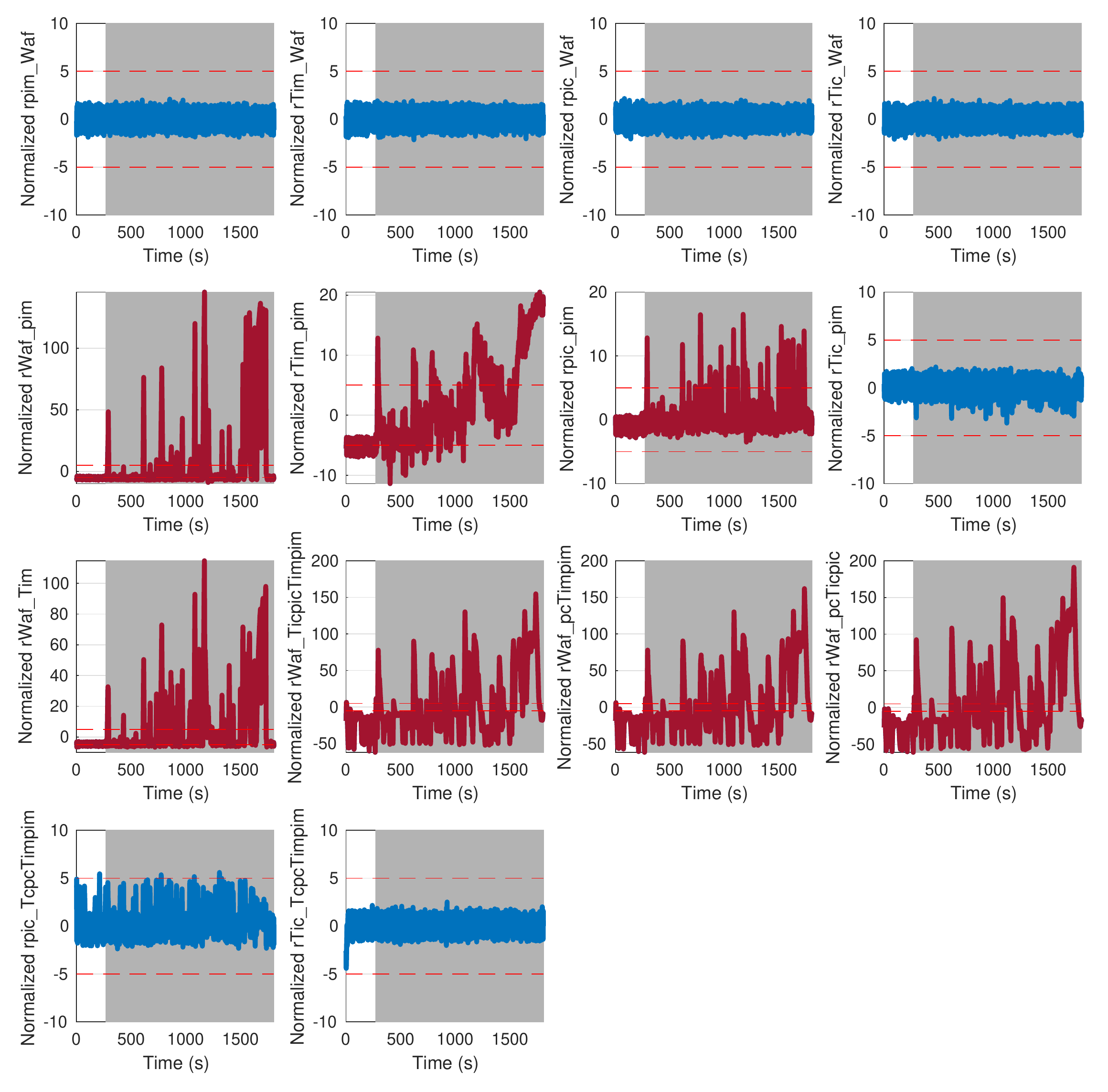}
  \caption{Normalized plots of the 14 additional residuals for a $f_{p_{af}}$ fault. The dashed lines are the fault detection thresholds. The plots in red are the residuals that have triggered. The shaded region shows the duration when the fault is active.}
  \label{fig:fpafadd}
\end{figure}

For the TCSI engine system, Algorithm \ref{alg:genRes} would generate the additional 441 potential candidates as well as their ideal FSM. Using these results, Algorithm \ref{alg:autoChoice} returned 14 best candidates as additional residuals that could be used to enhance fault isolability. The simulations were run again with the additional residuals and the shaded rows in Table \ref{tab:FSMdef} show their sensitivity towards the faults, i.e. if they have triggered or otherwise (Figs. \ref{fig:faultfreeadd} and \ref{fig:fpafadd} show some simulation results).

\begin{figure}[t!]
  \centering
  \includegraphics[width=0.8\columnwidth]{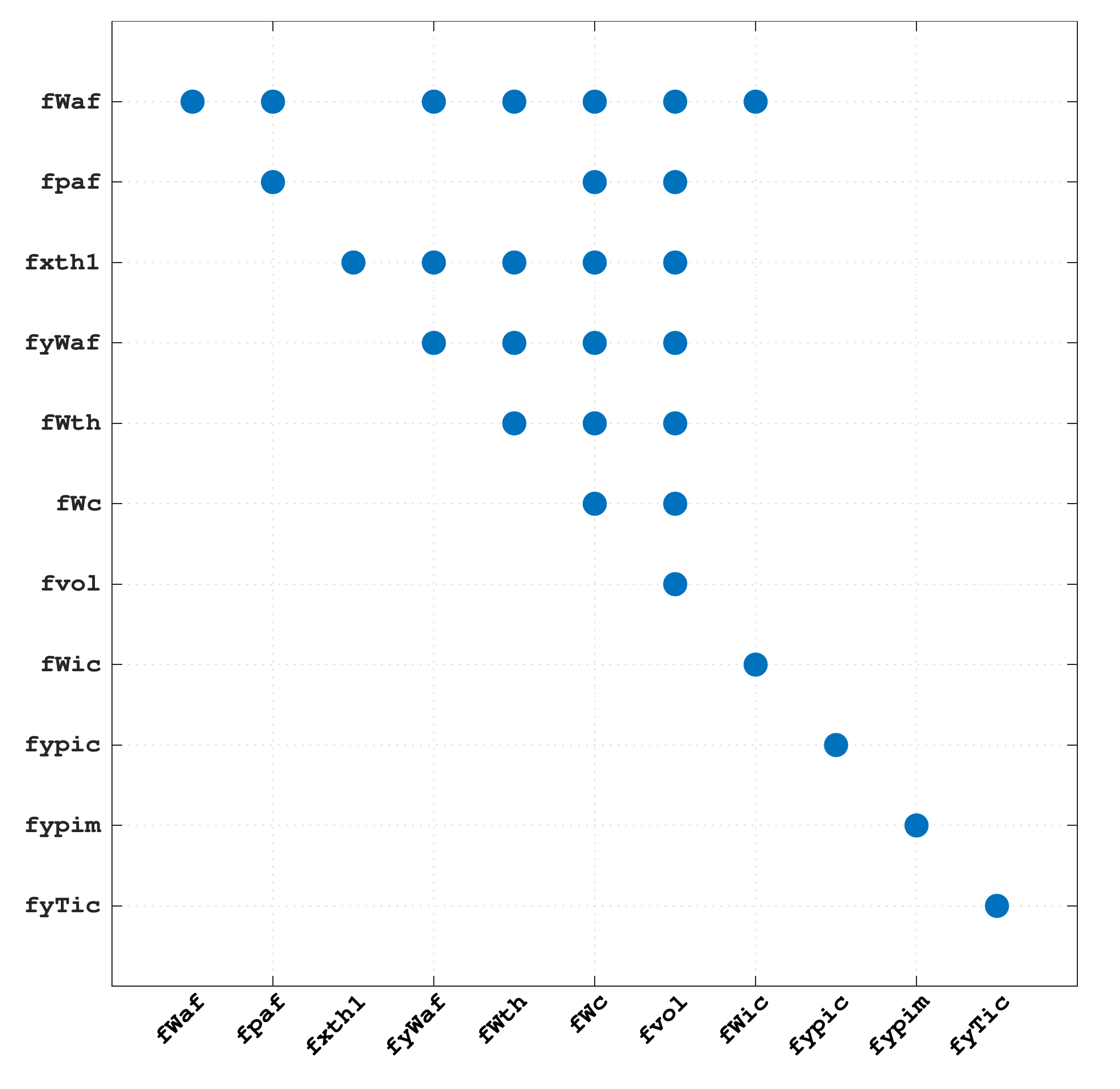}
  \caption{The FIM for the ``Original 7'' and the 14 additional residuals in Table \ref{tab:FSMdef} (all rows including the shaded rows).}
  \label{fig:FIMadd}
\end{figure}

Computing the FIM for the FSM of the original residuals together with this new set of best additional candidates produced a much desirable result as shown in Fig. \ref{fig:FIMadd}, where better fault isolation could be achieved compared to Fig. \ref{fig:FIMdef}. With the additional residuals, five faults: $f_{W_{af}}, f_{x_{th}}, f_{y_{T_{ic}}}, f_{y_{p_{ic}}}$, and $f_{y_{p_{im}}}$ can now be isolated from the other faults.

Furthermore, the FIM also shows that the detection of any actual fault involves less possible detection of other faults as potential candidates compared to Fig. \ref{fig:FIMdef}, hence improving the overall fault isolability.

\section{CONCLUSION} \label{sec:Conclusion}
This paper has presented the application of using additional residuals to enhance fault isolation on a vehicular engine system. The algorithm used to select a minimum number of best candidates for additional residuals helps to reduce the computation requirements to process the residuals, which allows for diagnosis to be performed onboard the vehicle in addition to cloud-based diagnosis architectures that many automakers are adopting nowadays. Also, by using structural analysis method, it has been shown that the presented method is able to improve fault isolability without adding physical sensors to the system. This would enable automakers to reduce hardware-redundancy in the design of the engine system, hence reducing the size and manufacturing cost of the system while ensuring safety and reliability. Future developments of this research include but not limited to further improving fault isolability using fault ranking, a combination of model-based and data-driven methods, or hybrid/hierarchical machine learning algorithms.

\addtolength{\textheight}{-5cm}   



%

%

\bibliographystyle{IEEEtran}
\bibliography{IEEEfull,refs}

\end{document}